# On/off switching of bit readout in bias-enhanced tunnel magneto-Seebeck effect

Alexander Boehnke, Karsten Rott, and Günter Reiss

Center for Spinelectronic Materials and Devices, Bielefeld University, Universitätsstrasse 25, Bielefeld, Germany

Andy Thomas

Thin films and Physics of Nanostructures, Bielefeld University, Universitätsstrasse 25, Bielefeld, Germany and Institut für Physik, Johannes Gutenberg Universität Mainz, Staudingerweg 7, Mainz, Germany

Christian Franz, Michael Czerner, and Christian Heiliger

I. Physikalisches Institut, Justus-Liebig-Universität Gießen, Heinrich-Buff-Ring 16, Gießen, Germany

Marius Milnikel, Marvin Walter, Vladyslav Zbarsky, and Markus Münzenberg

I. Physikalisches Institut, Georg-August-Universität Göttingen, Friedrich-Hund-Platz 1, Göttingen, Germany and Institut für Physik, Ernst-Moritz-Arndt Universität, Felix-Hausdorff-Str. 6, Greifswald, Germany

**Thermoelectric effects in magnetic tunnel junctions are currently an attractive research topic[1–3]. Here, we demonstrate that the tunnel magneto-Seebeck effect (TMS) in CoFeB/MgO/CoFeB tunnel junctions can be switched on to a logic "1" state and off to "0" by simply changing the magnetic state of the CoFeB electrodes. We enable this new functionality of magnetic tunnel junctions by combining a thermal gradient and an electric field. This new technique unveils the bias-enhanced tunnel magneto-Seebeck effect, which can serve as the basis for logic devices or memories in a "green" information technology with a pure thermal write[4] and read process[1–3]. Furthermore, the thermally generated voltages that are referred to as the Seebeck effect are well known to sensitively depend on the electronic structure and therefore have been valued in solid-state physics for nearly one hundred years. Here, we lift Seebeck's historic discovery from 1821 to a new level of current spintronics. Our results show that the signal crosses zero and can be adjusted by tuning a bias voltage that is applied between the electrodes of the junction; hence, the name of the effect is bias-enhanced tunnel magneto-Seebeck effect (bTMS). Via the spin- and energy-dependent transmission of electrons in the junction, the bTMS effect can be configured using the bias voltage with much higher control than the tunnel magnetoresistance and even completely suppressed for only one magnetic configuration, which is either parallel (P) or anti-parallel (AP). This option allows a readout contrast for the magnetic information of -3000% at**

**room temperature while maintaining a large signal for one magnetic orientation. This contrast is much larger than the value that can be obtained using the tunnel magnetoresistance effect. Moreover, our measurements are a step towards the experimental realization of high TMS ratios, which are predicted for specific Co-Fe compositions.**

A clear key-enabler for thermoelectric spintronic devices is a material system that exhibits large spincaloritronic effects and efficiencies or a high switching contrast between two thermoelectric states. For the Seebeck effect in magnetic tunnel junctions (MTJs), an important research target is the increase of the tunnel magneto-Seebeck (TMS) effect ratios. A promising route is to accordingly design the energy-dependent transmission function for electrons. In CoFeB/MgO/CoFeB tunnel junctions, this problem can be tackled because the $\Delta_1$ bands of the ferromagnetic CoFeB electrodes are highly spin-polarized, their position can be manipulated using the Co-Fe ratio[5] and the tunneling process is coherent[6]. Heiliger et al. recently calculated the TMS effect using *ab initio* methods for different Co-Fe compositions in MTJs with an MgO barrier[7]. They found that a change in the composition that alters the electronic states in the electrodes can be a powerful tool to tune both the junction's transmission function and the Seebeck coefficient[7]. The latter depends on the geometric center of the electronic occupation function. In a rigid-band model, this function can be designed by changing the electronic states on either side of the tunnel barrier[6]. In the rigid-band model, we increase or decrease the Fermi level by changing the Co-Fe composition (Fig. 1)[6]. However, the bias voltage $V_{bias}$ offers an alternative and experimentally much more accessible tool to shift the relative positions of the Fermi levels in MTJs. This option allows tailoring the energy- and spin-dependent tunneling properties, such as the tunnel magnetoresistance (TMR), which usually only decreases with increasing $V_{bias}$. The Seebeck effect and the TMS ratio, however, are expected to show a more complex behavior because the thermovoltage depends on the asymmetry of the electronic states with respect to the Fermi level[1]. Fig. 1 a illustrates, how the band shift induced by different Co-Fe compositions can vary the Seebeck effect from positive to negative. Here, we present two different experiments related with this approach. The first experiment introduces the bias-enhanced tunnel magneto-Seebeck (bTMS) effect, which exhibits a combination of currents that are generated by a voltage and a temperature gradient. As known from earlier investigations, the conductance and the Seebeck coefficients of an MTJ change differently when the magnetization is reversed[1,2]. Thus, the two currents can show an on/off behavior for specific bias voltages. Hence, the bTMS contrast ratio between the P and AP states can be much higher than for the TMR effect. Therefore, the bTMS is an ideal candidate for magnetic tunnel junction readout if a large signal is maintained for one magnetic configuration. The second experiment introduces a method to investigate the TMS effect under an applied bias voltage.

Up to now, it is not possible to totally switch off the conduction in a magnetic tunnel junction in one magnetic configuration, which is different from transistor devices, where the conduction channel can be blocked by the gate voltage. In this work, we attempt to overcome this drawback and increase the switching contrast, which is normally probed using the TMR when switching from the magnetically parallel (P) to the antiparallel (AP) configuration in an MTJ. For this purpose, we combine the TMS and TMR effects using simultaneously a thermal gradient $\Delta T$ and an effective field gradient $V$, to enable the control of the charge transport in each magnetic configuration by two degrees of freedom. From the Onsager transport equation for coherent transport, we obtain the contributions of both gradients to the total charge current[8–11]. Here, we compare the currents $I_{\text{on/off}}$ that are driven through the MTJ when it is heated by the laser and when the laser is switched off:

$$\Delta I_{P,AP} = I_{\text{on}} - I_{\text{off}} = \frac{S_{P,AP}}{R_{P,AP} - \Delta R_{P,AP}} \Delta T + \frac{1}{R_{P,AP} - \Delta R_{P,AP}} V - \frac{1}{R_{P,AP}} V \quad (1)$$

$$= \frac{1}{R_{P,AP} - \Delta R_{P,AP}} \left( S_{P,AP} \Delta T + \frac{\Delta R_{P,AP}}{R_{P,AP}} V \right)$$

$R$ is the resistance of the non-heated MTJ, and $R - \Delta R$ is the resistance when the MTJ is heated[12]; the indices P/AP refer to the magnetic configuration. Obviously, $\Delta I$ consists of two parts depending on two gradients across the barrier: a current driven by $\Delta T$ that is proportional to the Seebeck coefficient $S$, and a voltage-induced current that is proportional to the resistance ratio $\Delta R/R$. In an MTJ, each of these three quantities $S, R$ and $\Delta R$ change when the magnetic state of the MTJ is reversed from parallel (P) to antiparallel (AP)[1,13]. In a bias voltage region where the contributions generated by $\Delta T$ and $V$ are comparable, this relationship allows to deliberately set the current $\Delta I$ in a combination of both $\Delta T$ and $V$.

Eq. (1) reveals two interesting experimental options to exploit the two independent driving forces of the current. In the first variation, we apply an external fixed voltage to the MTJ in the P state that cancels out the Seebeck voltage $S_P \Delta T$ to achieve $\Delta I_P = 0$. When then the magnetization is reversed to the AP state, $V$ remains constant, but the resistances will change to $R_{AP}$ and $\Delta R_{AP}$, and the Seebeck voltage will change its magnitude to $S_{AP} \Delta T$ because of the TMR and TMS effects. Accordingly, the measured $\Delta I_{AP}$ will differ from zero. Although, a change in $R$ and $\Delta R$ could compensate the change in $S$, this exact cancellation is extremely unlikely for CoFeB/MgO MTJs (the effect ratios for these effects differ by at least one order of magnitude, see supplementary information). Hence, we should receive an on/off switching of the measured current upon magnetization reversal. To quantify this effect, we define a bias-enhanced TMS (bTMS) ratio

$$\text{bTMS} = \frac{\Delta I_{AP} - \Delta I_P}{\min(|\Delta I_P|, |\Delta I_{AP}|)} \cdot (2)$$

In the second experimental variation, we estimate how the Seebeck coefficients $S_{P,AP}(V)$ change with the applied bias voltage. This experiment can provide first insights into the band structure effects on the TMS because we can carefully tune the relative position of the Fermi levels using the bias voltage $V$. To suppress non-linearities with respect to $V$, we choose a small voltage interval that exhibits a linear I/V characteristic, implying $R(V) = R = const.$ and $\Delta R(V) = \Delta R = const.$ Therefore, we can use a linear model to estimate the contribution of $(\Delta R/R) \cdot V$ to $\Delta I$. Using the information from the model, we can subtract the non-Seebeck contribution from the measured $\Delta I$ and determine $S\Delta T$, which allows to compute the TMS ratio[1–3,13]:

$$\text{TMS} = \frac{S_P - S_{AP}}{\min(|S_P|, |S_{AP}|)} \quad (3)$$

For the experiments, we use CoFeB(2.5 nm)/MgO(1.7 nm)/CoFeB(5.4 nm) pseudo spin valve structures. The elliptic MTJs with a size of 6 μm × 4 μm are heated using a modulated diode laser (modulation frequency 1.5 kHz, laser power 30 - 150 mW, diameter in focus ≈ 5 μm, wavelength 640 nm)[11], which creates an AC-Seebeck current. Simultaneously, a DC bias voltage of up to $V_{bias} = \pm 300$ mV misaligns the electrodes' Fermi levels[14] and creates an additional DC current, which is mostly independent of the heating. Therefore, we use a lock-in amplifier to only detect the AC component of the tunneling current $\Delta I$ (see Eq. (1)). An estimation using the typical Seebeck coefficients[1,2,11] results in negligible contributions of artifacts caused by, e.g., Peltier- and Thomson-effects in our samples (see supplementary information). The dependence of the resistance on temperature must be analyzed both in the P and the AP state. Therefore, we measured the differential conductance[15–17] $dI(V)/dV$ of the heated and non-heated MTJ in both magnetic states (Fig. 2 c) for the applied bias voltages between ± 20 mV. The results show that the conductance is constant in this voltage regime.

To evaluate the TMS and its dependence on $V_{bias}$, we first characterized the tunneling resistance of the MTJs. Fig. 2 a shows the dependence of the resistance on the magnetic field; the black arrows indicate the magnetizations of the ferromagnetic electrodes. The TMR, i.e., the relative change of the resistance upon magnetization reversal, which is defined as TMR=($R_{AP}$-$R_P$)/min(|$R_{AP}$|,|$R_P$|), attains approximately 150%, which is a typical value for MTJs with a thin MgO layer of good quality[14,18]. Accordingly, we can realize the readout of the MTJ's magnetic state by measuring the resistance. Nevertheless, the bias-enhanced TMS that is determined at a similar MTJ provides a much higher effect ratio of -3000% (Fig. 2 b), which makes the detection of the MTJ's state easier and more precise. In this particular case, the high ratio is realized by combining a bias voltage of -10 mV with a temperature gradient across the barrier, which is created using a laser power of 150 mW. The measured signal $\Delta I$ is the current difference between the heated and non-heated MTJ, which is approximately 0 nA in the P state and -2.5 nA in the AP state of the MTJ. The high effect ratio is created by this on/off behavior of the signal when the MTJ state is switched between P and AP. The high readout contrast and the on/off behavior are two advantages of the bTMS compared to the TMR effect when it is used to determine the state of an MTJ.

A more detailed investigation of this remarkable result is shown in Fig. 3 a, where the measured current difference $\Delta I$ is plotted as a function of the external magnetic field for different values of $V_{bias}$. For $V_{bias}$ between -20 mV and +20 mV, the measured values drastically vary and even change their sign (Fig. 3b), which again points out the striking result of our experiments: we can switch the signal $\Delta I$ on or off by reversing the magnetic state of the MTJ, which is also visible in Fig. 2b. Figs. 3 a-b show that $\Delta I$ reverses the sign from negative to positive for $V_{bias}$ between 0 mV and -10 mV. However, $\Delta I_P$ and $\Delta I_{AP}$ do not change their sign at the same bias voltage. We find a zero $\Delta I_P$ at -10 mV. This zero crossing of $\Delta I_P$ and the finite value of $\Delta I_{AP}$ yield an infinite bias-enhanced TMS effect according to Eq. (2). Using our experimental data, we obtained a bTMS ratio of approximately -3000% at $V_{bias}$ ≈ -10 mV as shown in Fig. 2 a and Figs. 3 a-c for a laser power of 150 mW. Moreover, we observe that $\Delta I_{AP}$ crosses zero at -3 mV, which is accompanied by a non-zero $\Delta I_P$. Accordingly, we also find an increased bTMS ratio for -3 mV. The divergences of the bTMS ratio become more obvious when we apply a linear model to the measured data and calculate the bTMS ratio (Fig. 3 c). This result and the zero crossings of $\Delta I_P$ and $\Delta I_{AP}$ were previously revealed in Eq. (1) and become evident by concentrating on small bias voltages in Fig. 3. Because a vanishing $\Delta I$ signal is only found in one state of the MTJ at a certain bias voltage, $\Delta I$ can be switched from zero to a finite value only by changing the magnetic state of the MTJ. This result implies a potentially infinite contrast, e.g., between the P state (logic "1") and the AP state ("0") if the readout is performed by the bTMS effect, which was introduced in this paper.

The second important experimental result allows a first investigation of the TMS effect's dependence on the relative position of the Fermi level of the electrodes. However, according to Eq. (1), the measured signal includes a component that linearly rises with V, as long as $\Delta R/R$ is constant with V. This correlation is valid for small bias voltages (-20 mV to 20 mV) as presented in Fig. 2 c. Hence, we can calculate the Seebeck voltage $S\Delta T$ by subtracting $(\Delta R/R) \cdot V$ from the measured $\Delta I$ shown in Fig. 3b. This determination is based on a linear model, which is adapted to the measured $\Delta I(V)$ curves to deduce $\Delta R$ as the only free parameter. This model considers the measured resistance $R$ and the current at zero bias $\Delta I(V = 0)$. Then, we extract the information on the variation of the Seebeck voltage $S\Delta T$ at small bias values (Fig. 3 d) based on Eq. (1).

We present measurements with different laser powers on a second similar MTJ in Fig. 4. The figure displays the dependence of $S\Delta T$ that is extracted from Eq. (1) on the bias voltage. These measurements show that a higher laser power results in a higher Seebeck voltage. Furthermore, we observe a nearly constant Seebeck voltage in the P state when the bias voltage is changed.

However, the Seebeck voltage in the AP state of the MTJ varies much more with the bias voltage, which causes a crossing of the P and AP voltages. For the 90 mW laser power, this crossing is observed at -15 mV and 5 mV bias, whereas at 150 mW, the crossing is observed at -9 mV and 5 mV. Hence, this technique is a step towards the determination of the bias-voltage-dependent Seebeck coefficients in MTJs.

An important task for future experiments is to find methods to increase the Seebeck contribution to the measured current signal. A precise determination of the Seebeck voltage can provide a deep insight into the transport phenomena such as magnon scattering[19] and may also pave the way to giant TMS ratios[20]. However, this task is challenging because the generation of a temperature gradient in an MTJ will unavoidably increase its base temperature. This increase in temperature will affect the resistance, which is required to determine the non-Seebeck contribution (cp. Eq. (1)). Vice versa, the Seebeck effect will always disturb the resistance measurement of the heated MTJ. An independent determination is only possible when the temperature dependence of the resistance is determined separately. This task is challenging because the temperatures of both electrodes (separated by only a 1 nm tunnel barrier) cannot yet be determined.

In conclusion, we presented the bias-enhanced tunnel magneto-Seebeck (bTMS) effect that drastically increases the magnetic readout contrast. Furthermore, we introduced a new technique to determine the Seebeck voltages under an applied bias by applying a linear model to the experimental data.

The bTMS effect combines two effective gradients across the barrier: a temperature gradient and a voltage. Both gradients drive charge currents that depend on different parameters. The bias voltage creates a current that mainly depends on the resistance. The temperature gradient generates a current that is additionally influenced by the Seebeck coefficient of the MTJ (Eq. (1)). Both parameters change differently when the magnetization of the ferromagnets is reversed, which causes an on/off behavior of the bTMS signal at specific bias voltages. Hence, the effect ratio diverges (Figs. 3 a-c), which allows a better readout contrast than the commonly used TMR effect. Experimentally, we observed values of nearly -3000 %.

In the second part of this paper, we demonstrated that the combination of the bTMS setup and the Onsager transport equations (Eq. (1)) can be used to cast light on the behavior of the TMS effect on the direct tuning of the tunneling probability. An external bias voltage is applied to the MTJ to alter the relative position of the Fermi levels of the two ferromagnetic electrodes. In the future, this technique can be compared to a Fermi-level tuning by different compositions (doping) of the ferromagnetic electrodes[5,21] and corresponding *ab initio* calculations[7].

Both techniques – Fermi-level tuning and bias-enhanced TMS – provide a large contrast of a physical property in the two magnetization states. This result makes them notably attractive for future applications of MTJs in logic devices and memories.

**Acknowledgements**

The authors gratefully acknowledge financial support by the Deutsche Forschungsgemeinschaft (DFG) in the priority program SpinCaT (TH 1399/4-1, MU 1780/8-1, RE 1052/24-1). A.T. is supported by the MIWF of the NRW state government.

**Methods**

**Tunnel junction fabrication** The layer stacks for the MTJs were produced in a UHV vacuum sputtering chamber (base pressure $10^{-9}$ mbar). The stacks were prepared on an MgO substrate and consist of Ta 10/$Co_{25}Fe_{55}B_{20}$ 2.5/MgO 1.7/$Co_{25}Fe_{55}B_{20}$ 5.4/Ta 5/Ru 3 (all thicknesses are given in nm). After the preparation, the stacks were annealed at 450°C for one hour in an external magnetic field of 300 mT. The MTJs were patterned using e-beam lithography and subsequent Ar ion milling. The Ta layer beneath the CoFeB serves as the bottom electrode.

SiO$_2$ or Si$_3$N$_4$ is placed adjacent to the MTJs as an insulator. In an additional lithography and etching step, Au bond pads were patterned on top of the MTJ as the top contact so that the top electrode of the MTJ remains optically accessible.

**Bias voltage Seebeck experiments** Fig. 1 shows the setup to measure an AC Seebeck current under the applied DC bias voltage in an MTJ that is heated using a modulated laser (modulation frequency = 1.5 kHz). The optical setup to create a temperature difference ΔT inside the MTJ is identical to that in Ref. 11. This gradient leads to an AC Seebeck current inside the MTJ, which is fed to a variable-gain transimpedance amplifier (Femto DLPCA-200 current to voltage converter, gain $10^5$ V/A to $10^6$ V/A, AC coupling). The cut-off frequency is at least 200 kHz with a rise time of 1.8 μs, which is sufficient for the measurements[11]. The amplifier is simultaneously used to apply the desired DC bias voltage of up to ±300 mV across the MTJ. For our measurement, the output signal of the transimpedance amplifier is fed to a lock-in amplifier set to AC voltage mode with an integration time of 100 ms. The reference signal for the lock-in amplifier is the signal of the waveform generator, which controls the laser modulation. Hence, the lock-in amplifier provides an improved signal to the noise ratio and rejects the DC current that is generated by the bias voltage. For comparison, TMR measurements at different bias voltages were performed with a Keithley 2400 source meter (see Ref. 11), which allows calculations of the Seebeck coefficients from the measured TMS current and the barrier resistance.

**Author contributions**

A.B. and M.W. set up the experiment; A.B., M.W., and M.Mi. performed the measurements and analyzed the data; V.Z., M.W., and K.R. prepared and characterized the TMR devices; A.T., M.Mü., and G.R. designed the research approach; A.B., M.W., M.C., C.F., A.T., C.H., M.Mü., and G.R. invented the model to determine the Seebeck voltages under the applied bias. A.B., A.T., M.W., M.Mü., and G.R. wrote the manuscript; A.T., C.H., M.Mü., G.R., and all other authors discussed the experiments and the manuscript.

**Additional information**

The authors declare no competing financial interest. Supplementary information accompanies this paper.

# Figures

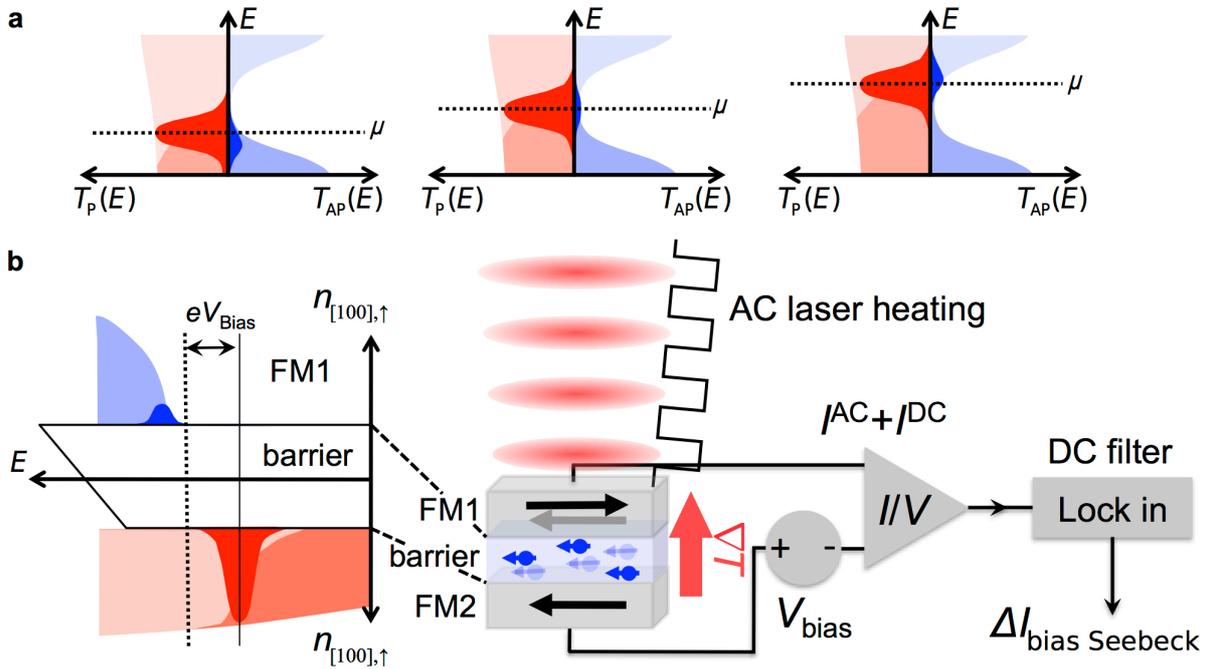

**FIG. 1 Shifting the electronic bands and the measurement setup. a** Exemplary transmission functions T(E) are plotted for the parallel (P) and antiparallel (AP) configurations; the derivative of the occupation function $\partial f(E, \mu, T)/\partial E$ is marked in dark color around the Fermi level. The three cases depict negative, zero and positive TMS values (asymmetry around the electrochemical potential µ). This result can be achieved by changing the electrode composition, and a comparable effect can be realized by tuning the bias voltage.

**b** Electrical setup with a transimpedance amplifier that converts the AC thermocurrent into an AC voltage, which is measured using the lock-in amplifier. The lock-in amplifier also serves as a filter for the DC signal that is generated by the bias voltage. The tunneling scheme depicts the transmission of the tunnel junction (density of states in [100] direction) in the antiparallel configuration, which is formed by the $\Delta_1$ bands and the barrier.

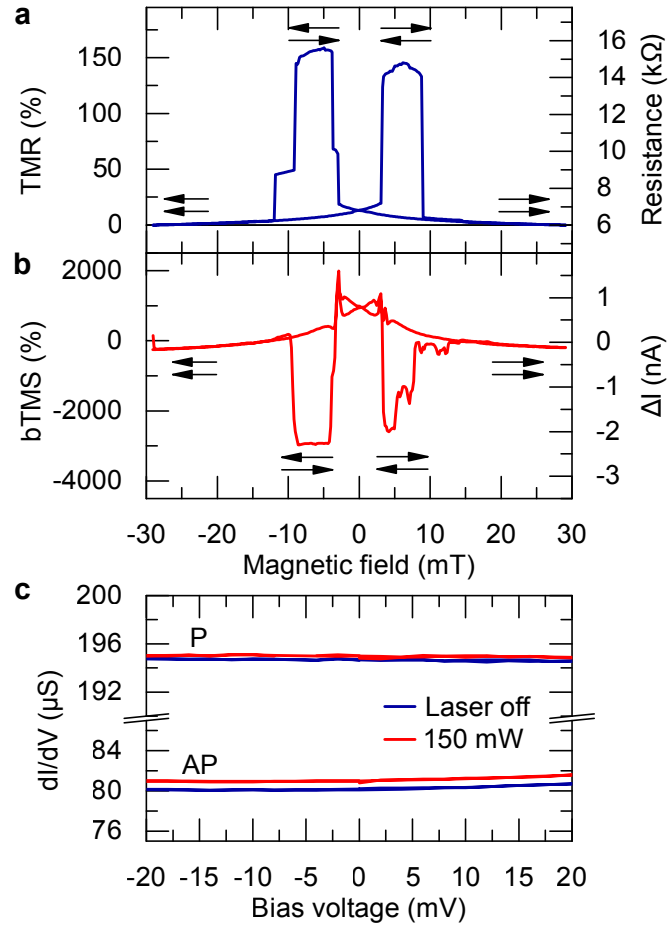

**FIG. 2 Tunnel magnetoresistance: a** TMR ratio and the resistance of the MTJ under a changing magnetic field. **b** Bias TMS ratio and measured the current signal for a bias voltage of -10 mV under 150 mW laser power. Here, the on/off behavior ($\Delta I_P = 0$ nA, $\Delta I_{AP} = -2.3$ nA) can be observed. Hence, the resulting effect ratio reaches nearly -3000 % and is much higher than the TMR ratio observed at the same MTJ. **c** Dependence of the differential conductance d$I$/d$V$ on the bias voltage for the heated (laser power 150 mW) and cold (laser blocked) MTJ. The values for the parallel (P) state and the antiparallel (AP) state have been measured at a magnetic field of 30 mT and -7 mT, respectively. In this small bias voltage range, the conductance is approximately constant.

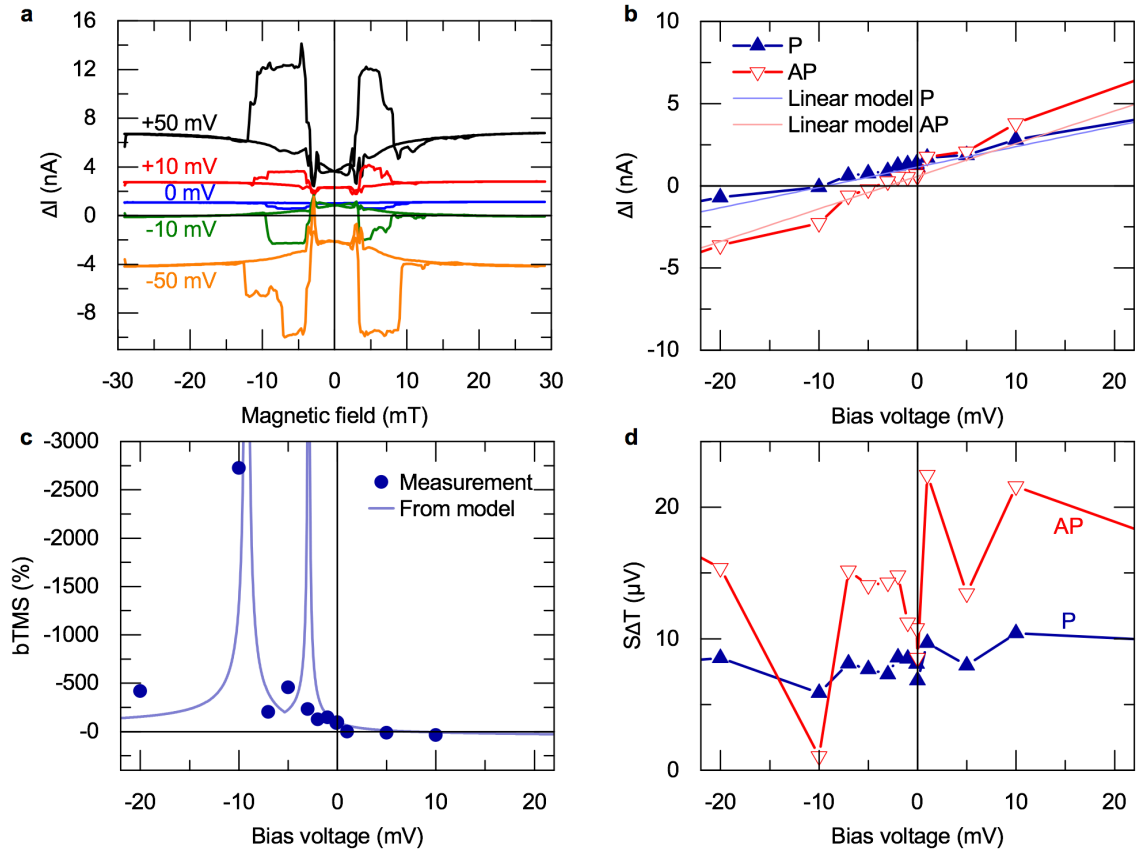

**FIG. 3 Bias tunnel magneto-Seebeck effect: a** ΔI versus magnetic field for selected bias voltages. At -10 mV, the P signal is close to zero, whereas the AP signal is non-zero. This difference produces an on/off behavior. **b**. Dependence of the measured current signal ΔI on the bias voltage for 150 mV laser voltage. $\Delta I_P$ and $\Delta I_{AP}$ cross zero at different values, which leads to an on/off behavior. The results from the linear model are shown as lines. **c** Bias TMS effect ratio derived from a. The light line is deduced from the linear model. The divergences and the high effect ratios are attributed to the vanishing ΔI in only one magnetic state of the MTJ (on/off behavior) at -10 mV and -2 mV bias voltage e. **d** Seebeck voltages that are derived from Eq. (1) after subtracting the linear contribution.

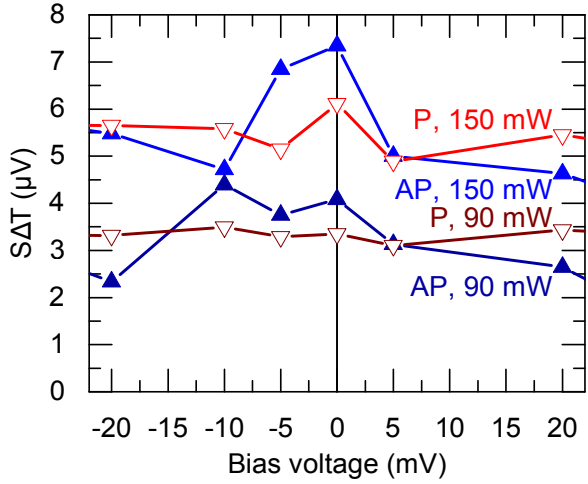

**FIG. 4 Dependence of SΔT on the bias voltage and heating power:** Seebeck voltages that are determined according to Eq. (1) for different laser powers in P and AP states. The signal rises with increasing laser power.

**TAB. 1 Resistance of hot and cold magnetic tunnel junction:** The resistances $R_{cold}$ and $R_{hot}$ were obtained from the dI/dV measurements using blocked laser and 150 mW laser power. These values can be used to separate the Seebeck and non-Seebeck components in Eq. (1).

| MTJ state | $R_{cold}$ (Ω) | $R_{hot}$ (Ω) | ΔR (Ω) | ΔR (% of $R_{cold}$) |
|---|---|---|---|---|
| P | 5136 | 5129 | 8 | 0.15 |
| AP | 12472 | 12342 | 130 | 1.05 |